# Wetting Transitions of Condensed Droplets on Superhydrophobic Surfaces with Two-Tier Roughness

Cunjing Lv, Pengfei Hao,[*] Xiwen Zhang and Feng He

Department of Engineering Mechanics, Tsinghua University, Beijing 100084, China

**Abstract**

**Although realizing wetting transitions of droplets spontaneously on solid rough surfaces is quite challenging, it is becoming a key research topic in many practical applications which require highly efficient removal of liquid. We report wetting transitions of condensed droplets occurring spontaneously on pillared surfaces with two-tier roughness owing to excellent superhydrophobicity. The phenomenon results from further decreased Laplace pressure on the top side of the individual droplet when its size becomes comparable to the scale of the micropillars, which leads to a surprising robust spontaneous wetting transition, from valleys to tops of the pillars. A simple scaling law is derived theoretically, which demonstrates that the critical size of the droplet is determined by the space of the micropillars. For this reason, highly efficient removal of water benefits greatly from smaller micropillar space. Furthermore, three wetting transition modes exist, in which the *in situ* wetting behaviors are in good agreement with our quantitative theoretical analysis.**

Understanding and realizing spontaneous wetting transition (for example, from the fully wetted Wenzel state to the partial wetted Cassie-Baxter state) of droplets on water-repellent materials is highly desired and of critical importance for a wide range of practical applications, such as self-cleaning, anti-icing, anti-corrosion, water harvest [1-4], and particularly in heat exchange technologies [5-7]. It is well known that dropwise condensation achieves heat and mass transfer coefficients over an order of magnitude higher than its filmwise counterpart [5,6] because small individual droplets can regularly form and shed off the surface before a thick liquid is formed, thereby minimizing the thermal resistance to heat transfer across the condensate layer. Unfortunately, even the most exciting natural superhydrophobic material, the lotus leaf, becomes sticky to condensed droplets [2], which strongly suppresses the probability of droplet wetting transition due to the pinning of the contact lines. For this reason, how to realize wetting transitions naturally and spontaneously on rough surfaces remains a key challenge, especially for small droplets. Despite that wetting behaviors of condensed droplets on superhydrophobic surfaces have been extensively studied [2,7,8-12], the wetting transition processes and our fundamental understanding of the underlying mechanism remains elusive to date.

---

[*] To whom correspondence should be addressed. E-mail : haopf@tsinghua.edu.cn

In this letter, by applying a thin layer of hydrophobic colloids to form two-tier rough surfaces, when the size (~ 10 μm in diameter) of the condensed droplet reaches a critical value, we report a wetting transition of a small droplet moving from the valley to the top of the micropillars autonomously, without any energy input or external force assistance. Three transition modes are captured, i.e., a single droplet, merging of two constrained droplets, and one constrained droplet triggered by some moving droplet. Based on systematic investigations, we show theoretically and numerically that further decreased Laplace pressure on the top side of the individual droplet accounts for such transition, and consequently allows the droplet to be squeezed out from the valleys. It is demonstrated that the critical size of the droplet for transition is governed by the space of the micropillars, irrespective of size. A general model is developed which is extremely effective in predicting the inner pressure of the droplet constrained in the pillars and its critical size for transition. We believe that these novel phenomena, as well as our fundamental understanding, shed new light on practical applications of superhydrophobic materials.

Silicon wafer substrate with square-shaped micropillars with side length $L$ and spacing $S$ of the neighbors are fabricated by photolithography and etching of inductively coupled plasma (ICP), and the height $H$ of the micropillars in this work is fixed at 5 μm. In order to obtain excellent superhydrophobicity, the substrates are subsequently treated with a commercial coating agent (Glaco Mirror Coat "Zero", Soft 99 Co.), in which nanoparticles are contained [13,14]. Surface topographies analyzed using a scanning electron microscopy (SEM, JSM 6330 from JEOL) are shown in Fig. 1. Figure 1 presents the details of the two-tier roughness due to the component of the self-assembled nanoparticles with a fractal-type structure. There are four types of samples investigated with $L$ = 2 μm, 3 μm, 4 μm, and 8.4 μm, respectively, and $S$ = 3 μm, 4.5 μm, 6.5 μm, and 13 μm, respectively. For a 5 μL droplet dropped on such substrates in an ambient environment, the average value of the apparent contact angles reaches as high as 161°, and the sliding angle is as low as 3.1° (see the Supplemental Material for additional details).

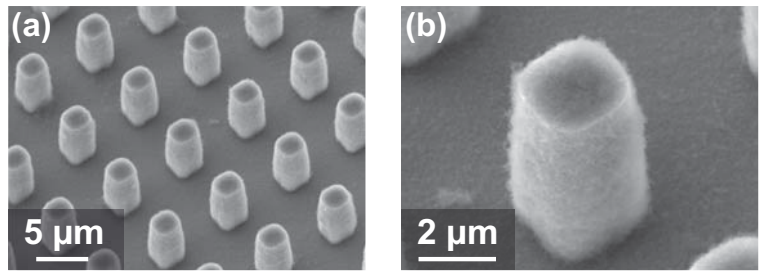

FIG. 1. Superhydrophobic surfaces ($L$ = 3 μm, $S$ = 4.5 μm, $H$ = 5 μm) with two-tier (micro and nano) roughness characterized by SEM in micro (a) and nanoscale (b), respectively.

Different from previous work typically carried out using the environmental scanning electron microscopy (ESEM) [9,15], we apply an optical microscopy technique under a moist ambient environment, which allows for focusing on the *in situ* dynamic characteristics of the condensed droplets, and overcomes the limitations of ESEM, such as slow imaging times, narrow ranges of operation pressures and temperatures, as well as beam heating effects [15]. All of the samples are placed horizontally on a peltier cooling stage, which is fixed on the slider of an optical microscope (BX51, Olympus, Japan). The optical microscopy that we employ allows *in situ* detection of the details with a high resolution of 0.2 μm. The laboratory temperature is measured at 29°C with a relative humidity of 40% (the corresponding dew point is 14°C). During the running of the cooling system, the temperature of the sample is well maintained at 10° ± 1°C. Top-down imaging of the processes are captured using a CCD camera (MegaPlus, RadLake, USA) installed on the microscope.

Figure 2 shows the dropwise condensation, wetting transition, and departure processes on one of the substrates in which $L = 3$ μm, $S = 4.5$ μm, and $H = 4.5$ μm. In the initial stage ($t < 60$ s), many isolated microdroplets are occurring on the side of the micropillars, as well as the valleys among the neighboring micropillars. In this stage, individual single droplets are growing and apparently spherical due to the excellent superhydrophobicity between water and the nanostructures [12-14]. Moreover, the size of the droplets remains smaller than $S$, which guarantees that coalescence will not occur.

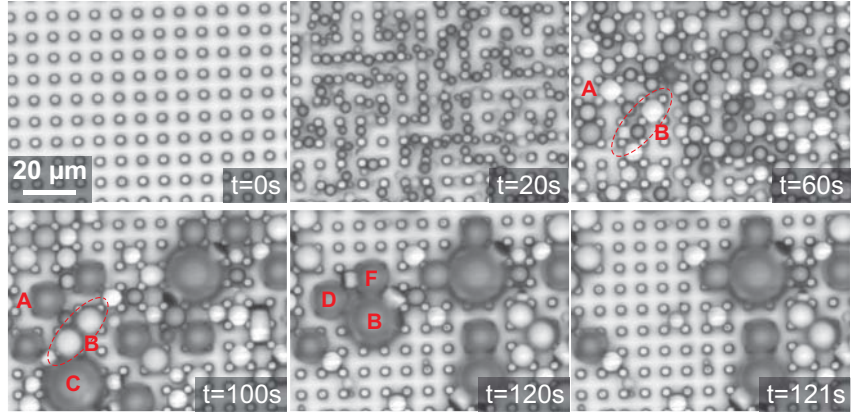

FIG. 2. Dropwise condensation on one of the two-tier substrates, which corresponds to $L = 3$ μm, $S = 4.5$ μm, and $H = 5$ μm. Droplets with a lighter color mean a direct contact on the substrate among the neighboring pillars. On the contrary, droplets with a darker color mean a non-contact with the substrate, i.e., such droplets attach to the side or on the top of the micropillars.

In the next stage ($t > 60$ s), when the size of the droplet reaches around 10 μm, we observe that some of the droplets unexpectedly climb to the top of the micropillars from the valley (e.g., one droplet marked as A from $t = 60$ s to $t = 100$ s, and two droplets marked B from $t = 100$ s to $t = 120$ s). In other words, the wetting transition occurs spontaneously without any external assistance, which is completely different from the partial Wenzel to Cassie transition [16], or external force induced dewetting (e.g., intense electric pulse [17], or vibration [18]). At the same time, coalescences occur and allow most of the droplets to merge with each other and jump continuously (e.g., droplet marked as C from $t$ = 100 s to $t$ = 120 s, droplets marked B, D, and F from $t$ = 120 s to $t$ = 121 s), which immediately leads to the formation of dry areas. Our observations reveal that generally, droplets experience the following three processes in this stage: (i) continuous growth of individual droplets; (ii) wetting transitions between different wetting states, i.e., from the valley to the top of the micropillars; and (iii) coalescence and departure. This letter will only focus on the second process.

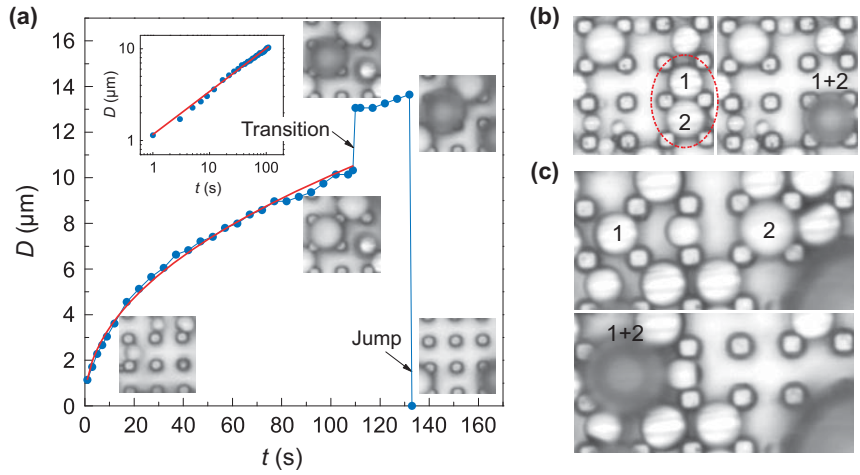

FIG. 3. Three types of wetting transition modes on the two-tier substrates ($L$ = 3 μm, $S$ = 4.5 μm, and $H$ = 5 μm). (a) *Mode – 1:* Time evolution of the diameter of an individual droplet during condensation (blue circles), in which the growth rate is scaled as $D \sim t^{0.47}$ (the red solid line). The insert covers $t \leq 109$ s in the log-log graph. The inserted individual frames along the growing processes correspond $t$ = 20 s, 109 s, 110 s, 133 s, and 134 s, respectively. There is a wetting transition at $t$ = 109 s and a jump at $t$ = 133 s. (b) *Mode – 2:* A wetting transition results from the merging of two isolated droplets marked as 1 and 2. (c) *Mode – 3:* A wetting transition results from a flying droplet from another location, i.e., droplet-2.

Unexpectedly, three wetting transition modes are observed. In all of them, the droplets could be spontaneously squeezed out of the valleys among the neighboring pillars, and the wetting behaviors are generalized below.

*Mode I: Wetting transition of an individual droplet.* — As shown in Fig. 3(a), condensation occurs on the substrate with $L$ = 3 μm, $S$ = 4.5 μm, and $H$ = 5 μm. Here, $t_0$ is defined at the moment when the tiny droplet is visible under the optical microscope. In the very beginning, the droplet is smaller than $S$, and hence it could only stay at the bottom of the valley or the side of the pillar ($t < 20$ s in Fig. 3(a)). When the size of the droplet exceeds $2^{1/2}S$, the gap between the neighboring pillars is filled, and the droplet is squeezed by the surrounding four pillars and deformed naturally away from a spherical shape (also see Fig. 2, droplets marked as A and B). While the droplets in the valleys continue to grow, a distinguishing phenomenon occurs: a wetting transition happens suddenly, and the droplet will stay on the top of the four pillars (from $t = 109$ s to $t = 110$ s in Fig. 3(a)), which is due to the upward motion of the contact lines along the micropillars. Furthermore, the apparent diameter of the droplet is 13 μm at $t = 110$ s, which is visibly larger than 10.3 μm at $t = 109$ s because of the constraint of the four pillars nearby. Fortunately, such a wetting transition could be judged by the distinct change of the contrast of the droplet. The lighter appearance of the droplet indicates a direct contact between the liquid and the substrate because the water is transparent and it could obtain better reflection from the substrate. On the contrary, the darker appearance of the droplet demonstrates that there is a gap between the substrate and the bottom of the lower side of the droplet, and hence the droplets just attach to the side or the top of the micropillars. Moreover, the diameter $D$ of individual droplets obeys a scaling law $D \sim t^{\alpha}$ [19,20], and $\alpha$ is determined as $\alpha \approx 0.47$ using the method of least squares. The deviation from the ~1/3 law may be due to the partial pinning of the contact line.

*Mode II: Wetting transition triggered by a static coalescence of two neighboring droplets.* — As shown in Fig. 3(b), initially, two neighboring droplets marked as 1 and 2 are growing and constrained inside of the micropillars (the lighter color of their appearance). However, they merge with each other when they become large enough, and the coalescence results in a wetting transition at the same moment (the darker color of the subsequent droplet). The remarkable feature of our observation is that at this moment there is only a static coalescence and a wetting transition is caused, which is quite different from the previous mobile coalescence phenomena [2], i.e., the out-of-plane jumping motion.

*Mode III: Wetting transition triggered by coalescence of a dynamic and a static droplet.* — As shown in Fig. 2, except for the wetting transitions on which we focus, mobile coalescence also occurs naturally, when the static droplet is triggered by a flying droplet (marked as 1 and 2 in Fig. 3(c), respectively), the static coalescence occurs again and results in a wetting transition.

This is the first report of static coalescence induced spontaneous wetting transition occurring on two-tier roughness surfaces, as well as three remarkable transition modes.

How do such interesting phenomena occur spontaneously? This question encourages us to more deeply investigate the variation of Laplace pressure $\Delta P = 2\kappa\gamma$ of the droplets with their size, in which $\gamma$ and $2\kappa = \kappa_1 + \kappa_2$ are the surface tension and the curvature of the liquid-vapor interface [21], respectively. Moreover, the curvature is constant for a certain volume because the gravity can be ignored due to the small size. $\kappa_1$ and $\kappa_2$ represent the two main curvatures displayed in Fig. 4(a). It is worth noting that there is no known analytical solution for the configuration of the droplet constrained and deformed by four micropillars (e.g., marked as A at $t = 60$ s in Fig. 2), in other words, $r \geq r_{min} = 2^{-1/2}S$ (Fig. 4(a)). Fortunately, we can pursue a first-order approximate solution (see the Supplemental Material for additional details):

$$\frac{\Delta P}{\gamma} = 2\kappa = -\frac{2\cos\theta}{S} + \left(4 + 2\sqrt{2}\cos\theta\right)\cdot\frac{1}{h}, \qquad (1)$$

where $\theta$ is the contact angle between water and the wall of the pillars; and $h$ is the height of the droplet. In Fig. 4(a)(b), the red circles and black squares represent numerical results (Surface Evolver [22]) of $\theta = 180°$ and $\theta = 150°$, respectively. The red and black solid lines are the corresponding results of Eq. (1), which can be utilized to predict $\Delta P$ perfectly with no fitting parameters. For convenience, all of the parameters are normalized by $\gamma$ and a length scale factor $a$ (here, we let $a = 1$ μm), i.e., $\overline{\Delta P} = \Delta P \cdot a/\gamma$, $\overline{\kappa} = \kappa \cdot a$ and $\overline{h} = h/a$, and we let $\overline{S} = S/a = 3$. If the droplet is too small and it just touches the edges of four pillars (which means that the droplet is a sphere with $r = r_{min}$), we directly have $2\kappa_{max} = 2/r_{min} = 2^{3/2}/S$. On the other hand, we can also get $2\kappa_{max}$ from Eq. (1) with case $h_{min} = 2r_{min}$, which is further verified in Fig. 4(b). On the contrary, if both $H$ and the volume of the droplet are big enough, we have $1/h \to 0$ and obtain $2\kappa_{\min} \to -2\cos\theta/S$ from Eq. (1). On the other hand, thus result can also be obtained directly and naturally by $\kappa_1$ ($\to -2\cos\theta/S$) because $\kappa_2$ disappears.

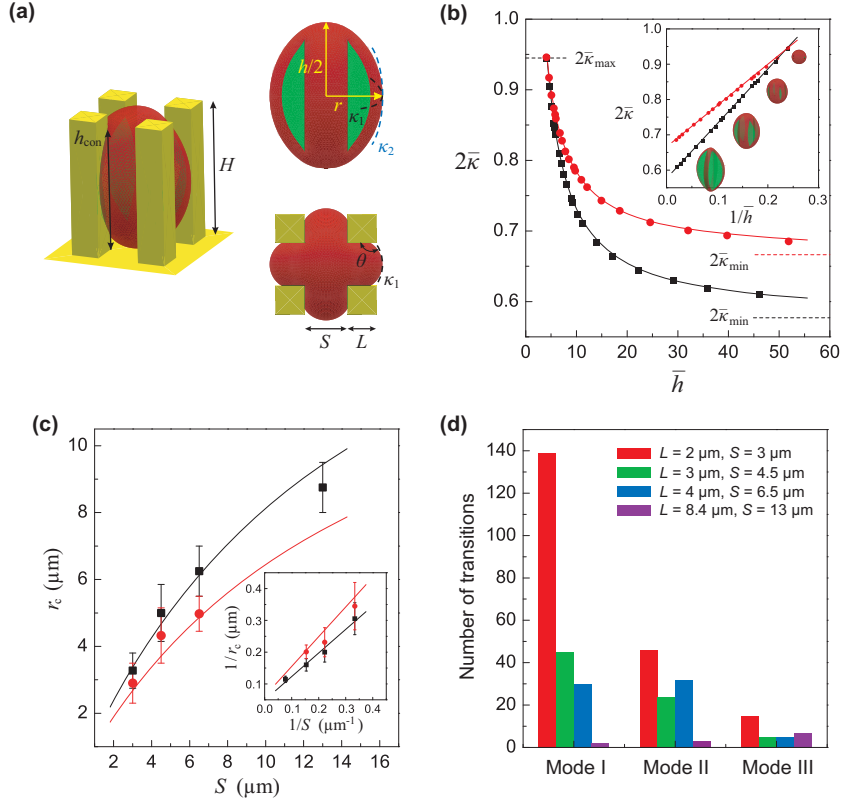

FIG. 4. (a) Definitions of the relevant geometrical parameters for a droplet constrained by four micropillars. $h_{\text{con}}$ is the distance between the upper solid-liquid contact boundary and the substrate; $h$ is the height of the droplet; $r$ is half of the maximum width of the droplet; $\theta$ is the apparent contact angle between the liquid and the wall of the micropillars; and $\kappa_1$ and $\kappa_2$ are the two main curvatures. (b) Relationship between the normalized curvature and height of the droplet. Here, the normalized space is $S/a = 3$. The red circles and black squares represent $\theta = 180°$ and $\theta = 150°$, respectively. The red and black solid lines are the corresponding theoretical results of Eq. (1). (c) Relationship between the critical droplet radius $r_c$ and $S$. The black squares and red circles represent experimental measurements of mode I and mode II, and the black and red solid lines are predictions using Eq. (3). (d) Number of wetting transitions vs transition modes. All of the statistics are carried out on a 350 μm × 114 μm area for a condensation duration of 600 s.

When condensation continues until $h_{\text{con}} \to H$, the lower side of the droplet is still constrained by the four micropillars. However, in the upper side, the solid-liquid contact lines start to spread to the top of these pillars, which results in the expansion of the droplet in the upper side. In other words, a pressure difference inside of the droplet will be created, i.e., $\Delta P_{\text{upper}} < \Delta P_{\text{lower}}$, so the liquid tends to move upward and the droplet starts to escape from the constraint of the four pillars. So, $h_{con} = H$ is treated as the critical condition to judge when the wetting transition occurs. At this moment, the other geometrical parameters also approach critical values, i.e., $h = h_c$, $r = r_c$, and $2\kappa = 2\kappa_c$ (or $\Delta P = \Delta P_c$). Here, $r_c$ is defined as the critical radius of the droplet just before the moment of transition (e.g., $t = 109$ s in Fig. 3(a)). Thus, we have $2\kappa_c = -2\cos\theta/S + (4 + 2^{3/2}\cos\theta)/h_c$. On the other hand, we define $2\kappa_c = 2/R^*$, where $R^*$ is employed to characterize the critical size of the droplet when transition happens. Thus, we have:

$$\frac{2}{R^*} = -\frac{2\cos\theta}{S} + \left(4 + 2\sqrt{2}\cos\theta\right) \cdot \frac{1}{h_c}, \tag{2}$$

Let $\beta = R^*/r_c$, $\eta = h_c/H$ be two geometrical coefficients. Then, we rewrite Eq. (2) to:

$$\frac{1}{r_c} = -(\beta\cos\theta) \cdot \frac{1}{S} + \frac{\beta}{\eta} \cdot \left(2 + \sqrt{2}\cos\theta\right) \cdot \frac{1}{H}, \tag{3}$$

As shown in Fig. 4(c), the black squares with error bars are the average values of five experimental measurements of the wetting transition of mode I. The black solid line is the result of Eq. (3) using the method of least squares, $1/r_c = 0.73/S + 0.24/H$. Without loss of generality, let $\theta = 160°$. Then, we get $\beta = 0.78$ and $\eta = 2.2$. We attribute the large value of $\eta$ to contact angle hysteresis and pinning of contact lines in a real moist ambient environment which results in larger deformation of the droplet. Let discuss the validity of $\beta$. On one hand, based on Eq. (2) we know $R^* \varepsilon [S/2^{1/2}, -S/\cos\theta] \approx [0.7S, S]$, on the other hand, we get $1/r_c = 0.73/S + 0.24/H$ ($H = 5$ μm) directly from the experimental data as mentioned. Thus, we roughly estimate that $R^*/r_c \varepsilon [0.5, 0.75] + 0.2\ S/H$, so the coefficient obtained ($\beta = 0.78$) is covered by the estimation, which further indicates our theoretical analysis and experiments are self-consistent. Furthermore, for mode II, when two droplets with $R^*_1$ and $R^*_2$ ($R^*_1 \approx R^*_2$) merge with each other, we will have a new droplet with radius $R^* \approx 2^{1/3}R^*_1$. Thus, naturally we have $r_{c\ (mode\ II)} = 2^{1/3} r_{c\ (mode\ I)}$ (plotted as the red solid line in Fig. 4(c)), which agrees well with the experimental results.

In Fig. 4(d), we present the statistical results between the number of wetting transitions and transition modes, as well as the influence of the space of the micropillars. All of them are carried out on a 350 μm × 114 μm area in 600 s. We can see that decreasing the space of the micropillars is a good strategy to decrease the critical size of the droplet to better realize dropwise condensation and suppress its filmwise counterpart. In this way, thermal resistance could be significantly reduced, and heat transfer enhancement will benefit greatly from this strategy. Actually, the energy barrier for the transition is scaled as $\Delta E \sim r_c^2 \gamma$, and the wetting transition benefits greatly from smaller $r_c$ (i.e., smaller $\Delta E$). This is the reason why the probabilities of the number of transitions are so dominant with the smallest value of space.

In conclusion, we report spontaneous wetting transitions of condensed microdroplets (~ 10 μm) on superhydrophobic substrates owing to two-tier roughness, and three wetting transition modes are observed. We not only present the theoretical expression which can be utilized to predict the Laplace pressure of the droplet with no fitting parameters, but also construct an explicit model which links the critical size of the droplet, and the size and height of the micropillars together. Our experimental and theoretical results indicate that further decreased value of Laplace pressure on the top side of the individual droplet leads to instability, and subsequently a surprising spontaneous wetting transition without any external force. We further reveal that the space of the micropillars is essential for determining the critical size of the droplet for transition. Thus, the design of devices with highly efficient removal of water in real applications can be inspired by this idea.

This work was supported by the National Natural Science Foundation of China (Grant Nos. 11072126, 111721156).

# Supplementary Material for

# "Wetting Transitions of Condensed Droplets on Superhydrophobic Surfaces with Two-Tier Roughness"

Cunjing Lv, Pengfei Hao,[*] Xiwen Zhang and Feng He

Department of Engineering Mechanics, Tsinghua University, Beijing 100084, China

## S1. Substrate fabrication and wetting characterization

The microstructured surfaces on the silicon wafer are fabricated by photolithography and etching of inductively coupled plasma (ICP). Then they are produced by treatment with a commercial coating agent (Glaco Mirror Coat "Zero", Soft 99 Co.) containing nanoparticles and organic reagent [1,2]. The superhydrophobic coating was applied on the smooth silicon wafer by pouring the Glaco liquid over the substrate. A thin liquid film wets the silicon surface and dries in less than one minute. Then the silicon surfaces are put into an oven and kept at 200°C for half an hour. The pouring and heating processes will be performed three to four times. Surface topographies are analyzed using a scanning electron microscopy (SEM, JSM 6330 from JEOL). Topology of harvested micro-nanostructured pillars with square shape are shown in Fig. 1, in which $L$ is the side length of the pillar, $S$ is the spacing between the nearest neighbors. $H$ is the height of the pillar and fixed at 5 μm. Fig. 1(a)(b) clearly demonstrate that the coating is composed of self-assembled nanoparticles with a fractal-type structure. There are 4 types of samples are investigated in our experiments, and their geometrical parameters are listed in Table S1, as well as the wetting properties. $\theta_a$, $\theta_r$, $\Delta\theta$, $\theta^*$ and $\alpha$ refer to the advancing contact angle, receding contact angle, contact angle hysteresis, static apparent contact angle and sliding angle of a macroscopic sessile droplet (5 μL) performed on these dry substrates. The small value of contact angle hysteresis and sliding angle indicate that these substrates with two-tier roughness own excellent superhydrophobicity. During the running of the cooling system, three types of wetting transition modes happen spontaneously and naturally as shown in Fig. S1.

---

[*] To whom correspondence should be addressed. E-mail : haopf@tsinghua.edu.cn

**TABLE S1:** Geometrical parameters and wetting properties (a 5 μL droplet) of the samples.

| No. | $L$ (μm) | $S$ (μm) | $\theta_a$ (°) | $\theta_r$ (°) | $\Delta\theta$ (°) | $\theta^*$ (°) | $\alpha$ (°) |
|---|---|---|---|---|---|---|---|
| 1 | 2.0 | 3.0 | 168.9 | 152.5 | 16.4 | 163.2 | 1.9 |
| 2 | 3.0 | 4.5 | 164.4 | 148.9 | 15.5 | 159.6 | 3.1 |
| 3 | 4.0 | 6.5 | 165.8 | 148.7 | 17.1 | 161.7 | 2.2 |
| 4 | 8.4 | 13.0 | 160.0 | 149.5 | 10.5 | 159.9 | 2.3 |

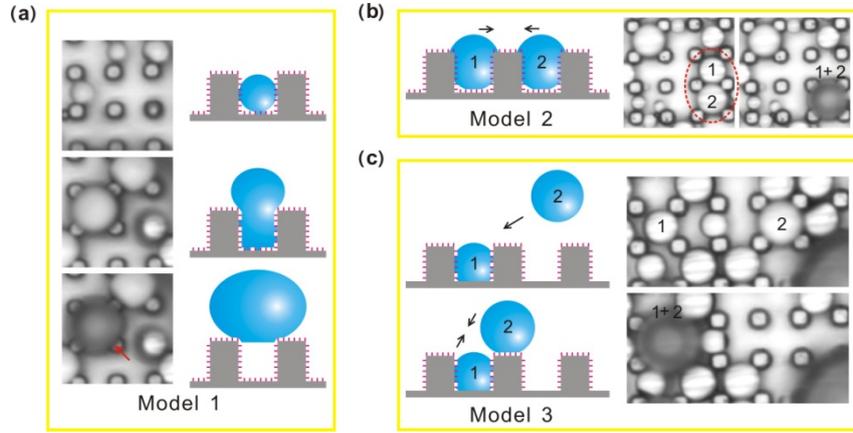

FIG. S1. Three types of wetting transition modes on the two-tier substrates ($L = 3$ μm, $S = 4.5$ μm, and $H = 5$ μm) observed experimentally and displayed schematically with more details.

## S2. Numerical simulations for the wetting transition behaviors
### S2.1. Geometry and Laplace pressure

For droplet with complicated geometry and in the absence of analytical solutions, we resort to a Finite Element Method (FEM). One public domain software package called SURFACE EVOLVER (SE) [3], developed by K. Brakke, is used to simulate the evolution of droplets during condensation. The basic concept of SE is to minimize the energy and the equilibrium shape of interfaces subject to the user defined surface tensions, interface areas, and constraints. SE can handle an arbitrary topology of the liquid surface, and during the calculations, it can report total energies, surface areas of the interface, and many other variable quantities.

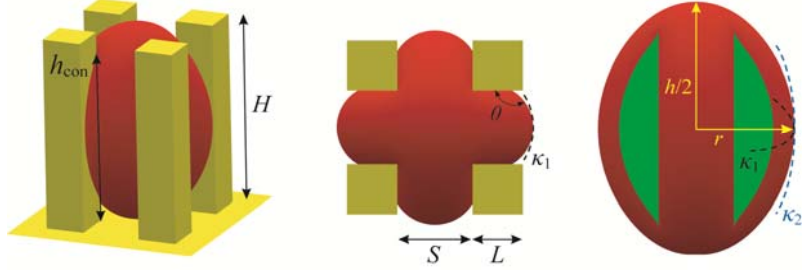

FIG. S2. Geometry of the droplet inside of four micropillars is displayed. $H$, $S$, $L$ are the height, space and with of the micropillars. $h_{\text{con}}$ is the maximum position of the solid-liquid contact boundary line. $\theta$ is the apparent contact angle between water and the wall of the micropillars. $\kappa_1$ and $\kappa_2$ are the two main curvatures. $r$ is the maximum radius of the droplet and $h$ is the height of the droplet.

For convenience, we give dimensionless parameters. Because all the experimental phenomena performed in micro scale, we choose $a = 1$ μm as the length scale factor, together with surface tension $\gamma$, all the geometrical and physical parameters are normalized as,

$$\bar{L} = \frac{L}{a}, \quad \bar{A} = \frac{A}{a^2}, \quad \bar{V} = \frac{V}{a^3}, \quad \Delta\bar{P} = \frac{\Delta P}{\gamma/a}, \quad \bar{E} = \frac{E}{\gamma a^2}, \tag{S1}$$

where $L$, $A$, $V$, $\Delta P$ and $E$ represent length, area, volume, pressure and energy. All the relevant parameters are defined in Fig. S2.

In our simulations, the total surface energy $E$ is defined as $E = \gamma (A_{\text{LV}} - A_{\text{SL}} \cos\theta)$, in which $A_{\text{LV}}$ and $A_{\text{SL}}$ are the area of the liquid-vapor and solid-liquid phases. $\theta$ is the contact angle which is defined according to the famous Young's equation $\cos\theta = (\gamma_{\text{SV}} - \gamma_{\text{SL}})/\gamma$, in which $\gamma$, $\gamma_{\text{SL}}$ and $\gamma_{\text{SV}}$ are the liquid-vapor, solid-liquid and solid-vapor interfacial tensions. Here we have to emphasize that $\theta$ is the apparent contact angle between the liquid and the wall of the micropillars with nano roughness, in other words, we treat the nano roughness is flat. Without lose of generality, we will first give two examples: $\theta = 150°$ and $\theta = 180°$.

As shown in Fig. S3(b), we give the appearance of the droplet with different volumes. Here, we have to emphasize that we aim to find the general law between $\Delta P$ and the relevant geometrical parameters, so with increase of the volume of the droplet, $S$ is fixed, but $H$ and $L$ will increase correspondingly to constrain the droplet.

On the basis of differential geometry [4], there are two main curvatures $\kappa_1$ and $\kappa_2$

at any point of the liquid-vapor interface, and $2\kappa = \kappa_1 + \kappa_2 =$ const. for a certain volume without consideration of gravity. For a sandwich droplet [5-7], it has already known that the curvature $2\kappa_s$ could be well described by the following formula,

$$2\kappa_s = \frac{2}{s_s} + \frac{\pi}{2} \cdot \frac{1}{d_s}, \tag{S2}$$

in which $s_s$ is the distance between the two parallel plane where the sandwich droplet is constrained, $d_s$ is the diameter of the sandwich droplet in the circumferential direction [6,7].

In our case (Fig. S2), even though apparently the droplet is like a cross intersection of two sandwich droplets, unfortunately, there is no analytical solution because it deforms from a spherical shape. Inspired by Eq. (S2), we still can find some links between the curvature $2\kappa = \kappa_1 + \kappa_2$ and other geometrical parameters. For convenience, we choose the point at the equator of the droplet to make analysis. $\kappa_1$ can be linked to $S$ directly, i.e. $\kappa_1$ is scaled as $\cos\theta/S$ (because $(S/2)/\cos\theta$ approximates the radius of the meniscus constrained in the two facing walls of the micropillars). Here, we have to emphasize that in Fig. S2 $\kappa_1 \neq 1/r$, because $r$ represents the scale of the droplet, instead of the radius of the meniscus constrained in the two facing walls. In other words, the relationship between $\kappa_1$ and $r$ is still unknown. On the other hand, the cross section of the droplet in the vertical direction is like an ellipse, even though we could not obtain the exact value of $\kappa_2$ directly, it is scaled as $\kappa_2 \sim 1/h$ (because the curvature radius in this direction deforms from a spherical). So we express the curvature of the deformed droplet as,

$$2\kappa = -\frac{2\cos\theta}{S} + f(\theta) \cdot \frac{1}{h}, \tag{S3}$$

in which $f(\theta)$ is a coefficient, geometrically, it reflects the deformation of the constrained droplet far away from a spherical shape, and Eq. (S3) could be treated as a general expression.

Fortunately, we have a boundary condition: when the droplet is small enough and it just touches the four edges of the four micropillars, the droplet is a sphere and we have $2\kappa_{max} = 2/r_{min} = 2^{3/2}/S$ and $h_{min} = 2r_{min} = 2^{1/2}S$ no matter the value of the contact angle, which gives us,

$$\frac{2\sqrt{2}}{S} = -\frac{2\cos\theta}{S} + f(\theta)\cdot\frac{1}{\sqrt{2}S}, \quad (S4)$$

so we get $f(\theta) = 4 + 2^{3/2}\cos\theta$, and finally we get Eq. (1),

$$\frac{\Delta P}{\gamma} = 2\kappa = -\frac{2\cos\theta}{S} + \left(4 + 2\sqrt{2}\cos\theta\right)\cdot\frac{1}{h}. \quad (1)$$

In Fig. S3, we give comparisons between numerical simulations (black squares and red circles) and Eq. (1) (black and red solid lines), here we let $S/a = 3$. Moreover, in Table S2, we also give the comparison between Eq. (1) and a direct fitting using the method of least squares (black and red dashed lines) with a formula $2\kappa = c_1/S + c_2/h$, here, $c_1$ and $c_2$ are two unknown coefficients.

Surprisingly, the theoretical results (Eq. (1)) are perfectly consistent with the numerical simulations. To the best of our knowledge, it is the first time we construct a direct relationship between the geometry and Laplace pressure of the constrained droplet. Eq. (S3) and Eq. (1) will significantly increase our understanding about the underlying mechanism for condensed droplet on microstructured surfaces.

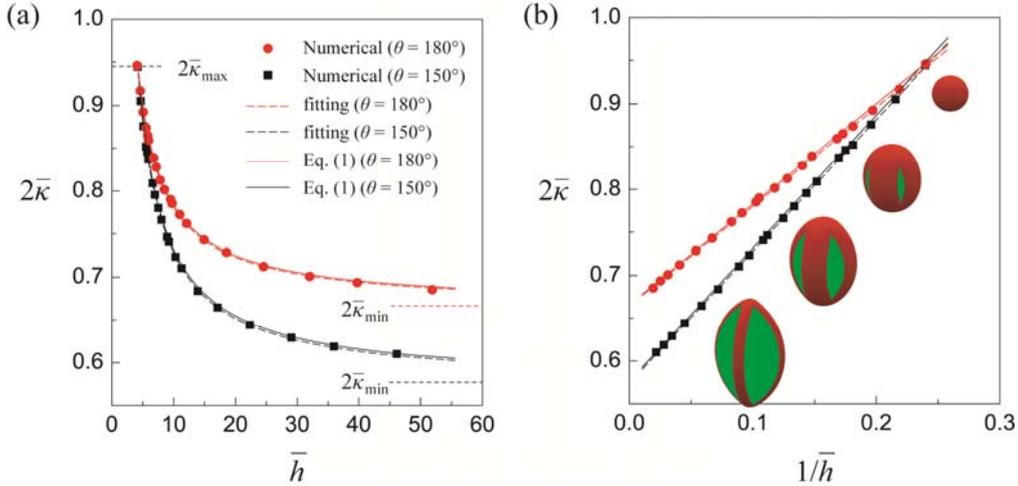

FIG. S3. Relationship between the height and Laplace pressure of droplets in different plots (a) and (b). The red circles and black squares are numerical simulations using Surface Evolver. The red and black solid lines are theoretical results of Eq. (1). The red and black dashed lines are theoretical results basis on fittings using the method of least squares with a formula $2\kappa = c_1/S + c_2/h$. Inserts in (b) represent droplets with different volumes, which correspond to $V/a^3 = 40, 100, 250, 2000$ from right top to left bottom.

**TABLE S2:** Comparisons between Eq. (1) and a direct fitting by the method of least squares.

| method | Eq. (1) | least squares method |
|---|---|---|
| formula | $2\kappa = -\dfrac{2\cos\theta}{S} + (4 + 2\sqrt{2}\cos\theta)\cdot\dfrac{1}{h}$ | $2\kappa = c_1\cdot\dfrac{1}{S} + c_2\cdot\dfrac{1}{h}$ |
| $\theta = 180°$ | $2\kappa = 0.6667 + 1.1716\dfrac{1}{h}$ | $2\kappa = 0.6655 + 1.1568\dfrac{1}{h}$ |
| $\theta = 150°$ | $2\kappa = 0.5774 + 1.5505\dfrac{1}{h}$ | $2\kappa = 0.5751 + 1.5370\dfrac{1}{h}$ |

## S2.2. Possible wetting states and transition

During condensation, small droplets appear and grow up. There would be some possible wetting states when the size of the droplet evolves. In Fig. S4, we demonstrate the relationship between the radius $\bar{r}_0 = r_0/a$ ($a$ = 1 μm) of one evolving droplet and its surface energy $\bar{E}$, here $r_0$ is used to defined the volume of the droplet $V_0 = 4\pi r_0^3/3$. In Fig. S4, the geometrical parameters are given as $\bar{L} = 2$, $\bar{S} = 3$ and $\bar{H} = 5$, which corresponds to the substrate No. 1 in Table S1. For convenience and without loss of generality, we let $\theta = 150°$ without any contact angle hysteresis or contact line pinning.

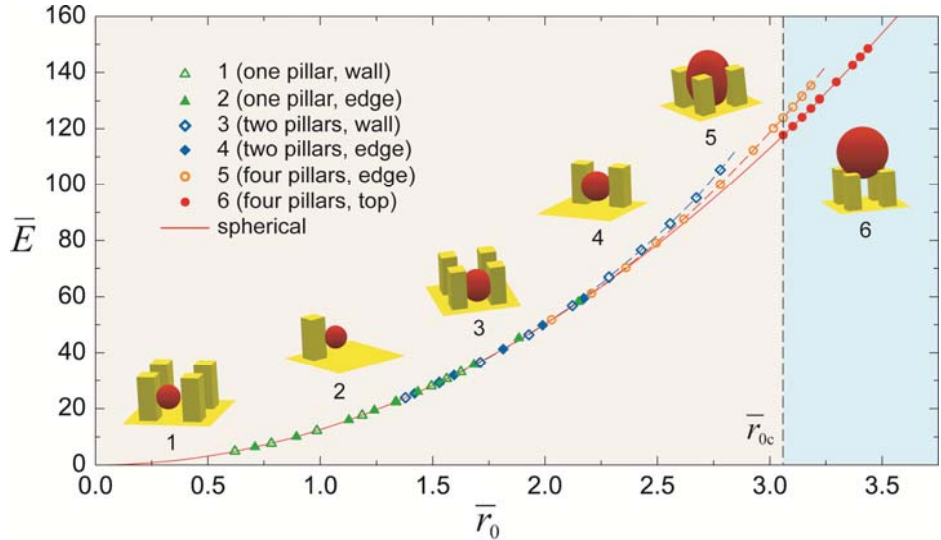

FIG. S4. Evolution of a single droplet during condensation, the relationship between the size of the droplet $r_0/a$ and the surface energy $E/(\gamma a^2)$ for different possible wetting states are given. The green and magenta dashed lines are used to guide eyes. The red solid line is a spherical droplet with $r_0/a$ aims to make a comparison.

When the droplet just appears, it's very small, so there are two possibly wetting state: "1" and "2" as shown in Fig. S4, in other words, the droplet could stay either on the wall of one individual micropillar, or stay at the edge of the pillar (for convenience, we only show one micropillar for "2"). When the droplet grows bigger, there will be another two wetting states: "3" and "4". For "3", initially the droplet performs like a sandwich droplet which is constrained by the parallel walls of two neighbor micropillars. Otherwise, the droplet will contact the edges of two micropillars, see "4". In wetting state "3", if the volume of the droplet grows big enough, surface energy will be accumulated and higher (which is demonstrated in Fig. S4, see the green hollow diamonds and dashed line), so there will be two possible ways to change the wetting states: the droplet will transfer from "3" to "4", or "3" to "5". If the size of the droplet increases further, it would touch three or four micropillars. We do not deny the probability of touching three micropillars exists, but due to the square shape of the micropillar alignment, the droplet will directly touch four micropillars (instead of touching three) when it becomes big enough, see wetting state "5". In order to make a comparison, we also plot the energy of a single droplet with spherical shape, i.e. the red line as shown in Fig. S4. We can see that the value of energy in wetting state "1-4" are very close to a spherical droplet with the same volume. However, in wetting state "5", because the droplet is totally constrained, increase the volume of the droplet will allow further increase of the surface energy (the magenta hollows circles and dashed line). After the size of the droplet increases much further, much more surface energy is accumulated and after it reaches a critical value, transition will happen, in other words, the droplet will go escape from the constraint of the pillars and will stay on the top of them, somehow like a Cassie-Baxter state. The energy barrier between "5" and "6" are clearly shown in Fig. S4, and the critical radius is found to be $\bar{r}_{0c} = 3.06$ (i.e., $r_{0c} = 3.06$ μm), which is consistent with our experimental results $r_c = 3.28 \pm 0.53$ μm (Fig. 4(c)) when $L = 2$ μm and $S = 3$ μm. Because of very high apparent contact angle, the wetting state of "6" is very close to a spherical droplet (the red solid circles and line).

**S1.3. Critical pressure**

In order to give a systematic investigation and understand more deeply, we try to extract more information about the droplet during wetting transition. Here, we want to emphasize that Surface Evolver is only available to deal with equilibrium or quasi-static problem, however, based on a careful control (Fig. S5), we still could capture the variations of its shape during transition, as well as the surface energy and Laplace pressure. In Fig. S5, we display the evolution at the critical moment when transitions happens (e.g. $r_0 = r_{0c}$ in Fig. S4).

In Fig. S5, $h_b$ is defined as the distance between the substrate and the bottom point of the droplet, $h(h_b)$, $r(h_b)$, $\Delta P(h_b)$ and $E(h_b)$ are functions of $h_b$. $\Delta P_0$ and $E_0$ are the Laplace pressure and the surface energy when the transition just starts to happen (i.e., the moment $h_b = 0$). $r_0$ is used to define the volume $V_0 = 4\pi r_0^3/3$ ($V_0$ is assumed to be constant in such a short wetting transition process).

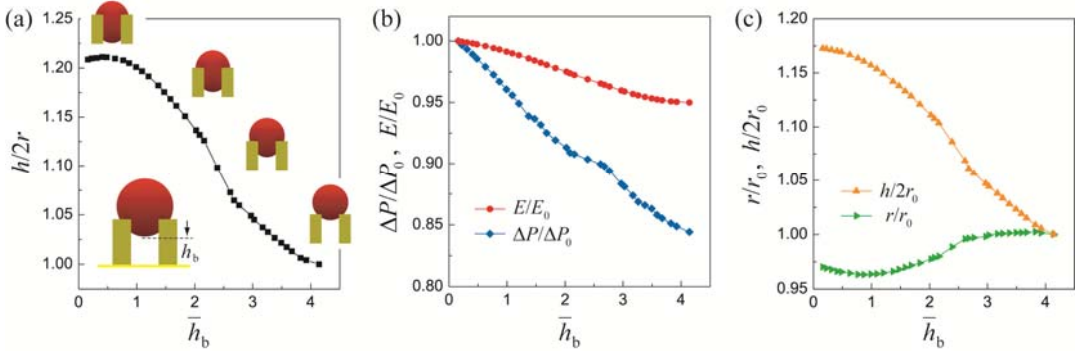

FIG. S5. Evolution of a single droplet at the moment of wetting transition.

In Fig. S5, just before the moment of transition, the droplet has an elliptical shape, $h/(2r) \approx 1.22$, but then it evolves to a spherical shape, which is displayed in Fig. S5(c). Fig. S5(b) tells us that the surface energy only decreases within 5%, however, the Laplace pressure decrease nearly 16% after the transition. These results suggest that because of excellent superhydrophobicity, deforming the droplet could not raise a huge barrier of surface energy, but arouse a remarkable Laplace pressure difference which is the main origin of the transition comes from.